\documentclass[11pt]{article}

\usepackage{amsmath,amssymb,amsthm,graphicx,latexsym}

\newcommand{\ed}{\end{document}}
\newcommand{\beq}{\begin{equation}}
\newcommand{\eeq}{\end{equation}}
\newcommand{\beqa}{\begin{eqnarray}}
\newcommand{\eeqa}{\end{eqnarray}}
\newcommand{\bc}{\begin{center}}
\newcommand{\ec}{\end{center}}

\newcommand{\ba}{\begin{array}}
\newcommand{\ea}{\end{array}}
\newcommand{\pa}{\partial}
\newcommand{\ta}{\theta}
\newcommand{\de}{\delta}
\newcommand{\brr}{({\bf r}-{\bf r}')}
\begin{document}

\title{{\bf{Noncommutative fluid and Growing Modes of Inhomogeneity in (Newtonian) Cosmology
 }}}
	\author{  {\bf {\normalsize Arpan Krishna Mitra$^1$}$
			$\thanks{E-mail: arpan@bose.res.in}}$~$
	{\bf {\normalsize Rabin Banerjee$^1$}$
			$\thanks{E-mail: rabin@bose.res.in}},$~$
		{\bf {\normalsize Subir Ghosh$^2$}$
			$\thanks{E-mail: subir\_ ghosh2@rediffmail.com}},$~$
		\\{\normalsize $^1$S. N. Bose National Centre for Basic Sciences,}
		\\{\normalsize JD Block, Sector III, Salt Lake, Kolkata-700098, India}
		\\\\
		{\normalsize $^2$Indian Statistical Institute}
		\\{\normalsize  203, Barrackpore Trunk Road, Kolkata 700108, India}
		\\[0.3cm]
	}
	\date{}
	\maketitle
	
\begin{abstract}
	Time evolution of modes of density contrast, in particular the growing modes, dictate the structure formation in Universe. In this paper we explicitly show how (spatial) Non-Commutativity (NC)  can affect the behavior of the modes, that is we compute NC corrected power law profiles of the density contrast modes. We develop a generalized fluid model that lives in NC space. The dynamical equations of fluid, namely the continuity and Euler equations receive NC contributions. When mapped to comoving coordinates these generate the NC extended versions of continuity and Friedmann equations for cosmology. Introducing cosmological perturbations finally yield the NC corrected evolution of density contrast modes. The construction of the NC fluid model from first principles and development of the formal aspects of its Hamiltonian formulation have been presented in the first part.	

\end{abstract}

	{\section{Intoduction}}

Noncommutative  (NC) spacetime effects in ideal fluid dynamics is turning into an area of recent activity. Our group is involved in a series of papers \cite{ar} concerned with generalized fluid models and various subtle aspects of Hamiltonian fluid dynamics. The present paper   in this series  concentrates on cosmological aspects of NC fluid dynamics. Other qualitatively distinct formulations of NC fluid dynamics appear in \cite{pra1,kai,van}.

Two distinct routes of introducing NC effects in fluid are the following: {\bf{(i)}} direct application of  the (Groenwald-Moyal) $∗$-product \cite{ncrev} in conventional Eulerian  field theory of fluid to introduce  NC-contributions \cite{van}.
{\bf{(ii)}} Introduction of NC algebra in  Lagrangian (discrete) fluid degrees of freedom  which subsequently percolates to the Euler (field) fluid degrees of freedom \cite{pra1,ar} and NC-extended fluid action \cite{pra1}.  

We have followed the second approach,
the primary reason being that NC generalization can be unambiguously done in the discrete variable setup. We stress that our model is constructed from first principles, completely based on the map
between the Lagrangian and Hamiltonian (Euler) formulation of fluid dynamics (see  \cite{jac} for a detailed discussion of the formalism for canonical fluid). The NC effect is induced
in the Euler continuum algebra from the NC extension in discrete Lagrangian variable algebra.

In this perspective a few comments about the role of fluid dynamics in topical theoretical physics would be worthwhile.  Hydrodynamics can be described in very general terms as a universal
description of long wavelength physics that deals with low energy effective degrees of freedom of a field
theory, classical or quantum. Interestingly it applies equally well at macroscopic and microscopic scales. The physical reasoning behind
the success of fluid models is the reasonable assumption that at sufficiently high energy densities local
equilibrium prevails in an interacting field theory so that local inhomogeneities of the discrete Lagrangian system are smoothed out to continuous fluid variables. Indeed the constitutive relations provide the bridge between
the fundamental(discrete) and the continuous degrees of freedom. In the fluid picture one simply deals with the age old
continuity equation and the Euler force equation, (in the most simple non-relativistic framework for an
ideal fluid). Two of the most exciting and topical areas of recent interest are cosmology and AdS-CFT
correspondence \cite{mal}  as applied to gauge-gravity duality \cite{gg}  and both of them rely heavily on conventional
fluid dynamics. Hence it will be very interesting to study the effects of non-trivial changes in fluid equations, brought about by introducing noncommutativity.
In the present paper we aim to outline how noncommutativity in fluid dynamics  can play an important role in  cosmology through cosmological perturbations.

A recently rediscovered effect that has impacted the theoretical physics community in a major way is
the idea of introducing a NC spacetime and subsequently a NC generalization of quantum mechanics and  field theory.
Seiberg and Witten \cite{ncrev}, in their seminal work, resurrected the (not so successful) NC spacetime introduced
by Snyder \cite{sny}, when the former demonstrated that in certain low energy limits open string dynamics
(with end points on $D$-branes) can be simulated by NC field theories where the NC parameter $\theta_{\mu\nu}$  is to be
identified with the anti-symmetric two-form field Bµν. This idea has led to a huge amount of literature
and we refer to a few review works in \cite{ncrev}. In the present paper we develop a generalized fluid dynamics,
compatible with spatial NC (since we work in non-relativistic regime), from first principles. One generally
tends to avoid introducing NC between time and spatial degrees of freedom due to possible complications
regarding unitarity in the quantum theory through introduction of higher order time derivatives.

Let us now come back to the classical scenario that is of present interest. Just as classical Poisson
algebra is elevated to quantum commutators via the correspondence principle, in exactly an identical fashion
one can think of a classical counterpart of the NC quantum algebra. Once again Jacobi identity, in the
sense of double brackets, plays an essential role since, for a Hamiltonian system the symplectic structure
(brackets) has to obey the Jacobi identity. However there is an added twist in the classical setup where
sometimes it is possible to judiciously introduce constraints in a particular model under study such that the
NC generalized algebra can be identified with the Dirac bracket algebra \cite{dirac}. However this line of approach is not exploited in the present work.

We outline our formalism following \cite{jac,ar,pra1}. The NC fluid model proposed by us rests essentially
on the map between the Lagrangian and Eulerian or (Hamiltonian) description of fluid dynamics. The
former is based on a microscopic picture where the fluid is treated as a collection of  a large number
of point particles obeying canonical Newtonian dynamics. The d.o.f.s consist of the particle coordinate
and velocity, $X_i(t), dX_i(t)/dt$, respectively, where $i$ stands for particle index. In the limit of a continuum, these reduce to $X(x, t), dX(x, t)/dt$ with $x$ replacing the discrete index $i$. On the other
hand the Eulerian scheme starts by providing a field theory Hamiltonian  and a set of
Poisson brackets between the fluid d.o.f that are the density and velocity fields
 $\rho (\bf r)$ and  ${\bf v}(\bf r)$
 respectively.
The fluid equations of motion are derived from the above as Hamilton’s equation of motion. The most relevant result from our perspective is that the Poisson brackets between Euler field variables are explicitly derivable from Poisson brackets between (discrete) Lagrangian d.o.f.. The chain of steps  leading from Lagrangian to Eulerian formulation is 
best suited for our purpose since, as discussed earlier, the NC brackets are given most naturally in point
mechanics framework, that is in terms of Lagrangian variables. It is worthwhile to recall here that even
the canonical point mechanics (Poisson) brackets lead to a quite involved and non-linear set of operatorial
algebra between the Euler variables. Hence it is not entirely surprising that the simplest extension of
canonical brackets to NC brackets in Lagrangian setup will lead to an involved and qualitatively distinct
NC extended brackets among Euler variables. However, as we will explicitly demonstrate, these NC brackets
yield a modified set of continuity equation and Euler force equation. Clearly these NC modifications will leave their marks
on cosmological solutions and in particular the NC corrections will act as specific forms of cosmological
perturbations.

The paper is organized as follows: in Section 2 we recapitulate the Hamiltonian or Euler form of fluid
dynamics along with its spacetime symmetries. Section 3 deals with our extension to NC fluid variable algebra with
a discussion on the corresponding Jacobi identities and a study of
the generalized continuity and conservation principles including comments on spacetime symmetries for
NC fluid. In Section 4 we provide an outline on the
 effects on cosmological principles induced by NC modified
fluid system.  In Section 5 we discuss in an explicit way the significance of our NC generalized cosmology. We conclude in Section 6 with a summary of our work and its future prospects.

\section{Hamiltonian formulation of Eulerian fluid: A brief review}

 Newton's law for the particle (Lagrangian) coordinate $X_i(t)$ and velocity $v_i(t)=\dot{X_i}$ is given by,
  \begin{equation}
 m\ddot 
  X_i(t)=m\dot v_i(t)=F_i(X(t)), \label{new}
  \end{equation}
 where $m$ is the mass of individual particle and $F_i(X(t))$ is the applied force. On the other hand the Eulerian density for the single particle is,
  \begin{equation}
  \rho (t, {\bf r}) = m  \delta ({\bf X} (t) - {\bf r}).
  \label{rs}
  \end{equation}
  For a number of particles the density  field is given by, 
  \begin{equation}
 \rho (t, {\bf r}) = m \sum^N_{n=1} \delta ({\bf X}_n (t) - {\bf r}).
 \label{r}
 \end{equation}
 It is straightforward to define the fluid current as,
 \begin{equation}
 {\bf j} (t, {\bf r}) = {\bf v} (t, {\bf r}) \rho (t, {\bf r}) = m \sum^N_{n=1} {\bf \dot{X}}_n (t) \delta ({\bf X}_n (t) - {\bf r}).
 \label{j}
 \end{equation}
 Finally  replacing the discrete particle labels  by continuous spatial  arguments (omitting time $t$) we arrive at,
 \begin{equation}
  \rho({\bf r})=\rho_0\int\delta(X(x)-r)dx,
  v_i({\bf r})=\frac{\int dx \dot{X_i}(x)\delta (X(x)-r)}{\int dx \delta(X(x)-r)}.
 \label{ncc13}
 \end{equation}
 The integration is over the entire relevant volume.
(The dimensionality of the measure will be specified only when formulas are dimension specific.) $\rho_{0}$
is a background mass density, so that the volume integral of density $\rho$ is the total mass.

  In a Hamiltonian formulation the canonical (equal time since we are in non-relativistic framework) Poisson bracket structure   is given by
  \begin{equation}
 \{\dot{X}^i , X^j \} = (i/m) \delta^{ij},~~\{X^i , X^j \} = 0,~~ \{\dot{X}^i , \dot{X}^j \} = 0.
 \label{nnp1}
 \end{equation}
 For the Lagrangian fluid this is generalized to   \cite{jac}, 
 \begin{equation}
 \{\dot{X}^i ({\bf x}), X^j ({\bf x'})\} = \frac{1}{\rho_0} \delta^{ij} \delta ({\bf x} -{\bf x'});~~ \{X^i ({\bf x}), X^j ({\bf x'})\} =  \{\dot{X}^i ({\bf x}), \dot{X}^j ({\bf x'})\} = 0.
 \label{p11}
 \end{equation}
 Obviously the above bracket structure satisfies the Jacobi identity.
Using the definitions of $\rho$ and $\bf j$ in terms of $\bf X$ and $\bf \dot{X}$ given above (\ref{ncc13}),  a straightforward  computation leads to the Poisson algebra between the Euler variables $\rho$ and ${\bf j}$ \cite{jac} (details of the computation are provided in the appendix): 
\begin{eqnarray}
\{\rho({\bf r}), \rho ({\bf r}')\} &=0& \label{1rj}\\
\{j^i ({\bf r}), \rho ({\bf r'})\}&=& \rho ({\bf r}) \partial_i \delta \brr \label{2rj}\\
\{j^{i} ({\bf r}),j^j ({\bf r}')\} &=& j^j ({\bf r}) \partial_i \delta \brr + j^i ({\bf r'}) \partial_j
\delta \brr . \label{rj}
\end{eqnarray}
Since ${\bf j}={\bf v} \rho$ an equivalent set of brackets follows \cite{jac}:
\begin{eqnarray}
\{v^i({\bf r}), \rho({\bf r'})\} = \partial_i \delta \brr, \qquad \label{rrv}\\
\{v^i({\bf r}), v^j({\bf r'})\} = -\frac{\omega_{ij} ({\bf r})}{\rho({\bf r})} \delta \brr,
\label{rv}
\end{eqnarray}
where
\begin{equation}
\omega_{ij} ({\bf r}) = \partial_i \ v_j {(\bf r)}-\partial_j v_i ({\bf r})
\label{v}
\end{equation}
is called the fluid vorticity, which vanishes for an irrotational fluid.

The Hamiltonian for a generic barotropic fluid (where pressure depends only on density) is taken as,
\begin{equation}
\label{hami}
H=\int dV~{\cal{H}}=\int dV~(\frac{1}{2}\rho v^{2} +V(\rho)).
\end{equation}
The pressure $P$ is related to $V$ via $P(\rho)=\rho \frac{\partial V}{\partial \rho}-V$. Fluid dynamical equations follow from the Hamiltonian equations of motion:
\begin{equation}
\label{2an1}
\dot{\rho}=\{\rho , H\}=-\pa_{i}(\rho v_{i}),
\end{equation}
\begin{equation}
\label{2al1}
\dot{v_{k}}=\{v_{k}, H\}=-v_{i}\pa_{i}v_{k} - \pa_{k}V'(\rho).
\end{equation}

\section{Non Commutative generalization}
 Let us now generalize the above to NC space. We start with the usual minimal (and most popular) form of extended NC Poisson brackets between the Lagrangian variables, 
  \cite{ncrev},
\begin{equation}
\label{aa}
\{X_{i}({\bf{x}}), X_{j}({\bf{y}})\} = \frac{\ta_{ij}}{\rho_{0}}\de ({\bf{x}}-{\bf{y}}),~
\{\dot{X_{i}}({\bf{x}}), X_{j}({\bf{y}})\}=\frac{1}{\rho_{0}}\de_{ij}\de({\bf{x}}-{\bf{y}}),~ \{\dot{X_{i}({\bf{x}})}, \dot{X_{j}}({\bf{y}})\}=0,
\end{equation}
where the NC parameter tensor $\theta_{ij}$ is  constant and antisymmetric $(\theta_{ij}=-\theta_{ji})$. This is the simplest extension of the canonical algebra (and qualitatively equivalent to the NC proposed by Seiberg and Witten in \cite{ncrev} to  NC space.

In Eulerian description in NC space we define the fluid variables in the same way as in \eqref{j} and \eqref{ncc13} and  the induced NC field algebra appears below (computational details are given in Appendix A1),
\begin{equation}
\label{ae}
\{\rho({\bf{r}}), \rho({\bf{r'}})\}
=-~\ta_{ij}\pa_{i}\rho \pa_{j}\de({\bf{r}}-{\bf{r'}}),
\end{equation}
\begin{equation}
\label{af}
\{\rho({\bf{r}}), j^{i}({\bf{r}})\}
=\rho({\bf{r'}})\pa_{i}\de({\bf{r}}-{\bf{r'}})-
\ta^{jk}\pa_{k}\de({\bf{r}}-{\bf{r'}})\pa_{j}j^{i}({\bf{r}}),
\end{equation}
\begin{equation}
\label{ag}
\{j_{i}({\bf{r}}), j_{j}({\bf{r}})\}= j_{i}({\bf{r'}})\pa_{k}\de({\bf{r}}-{\bf{r'}})+j_{k}({\bf{r}})\pa_{i}\de({\bf{r}}-{\bf{r'}})-\ta_{lm}\pa_{m}
\de({\bf{r}}-{\bf{r'}})\pa_{l}(\frac{j_{i}({\bf{r}})j_{k}({\bf{r}})}{\rho({\bf{r}})}).
\end{equation}
Again we have a similar set of equations between the density ($\rho$) and the fluid velocity ($v^{i}$).
\begin{equation}
\label{9y}
\{v_{i}({\bf{r}}), \rho({\bf{r}})\}=-\theta_{jk}\pa_{k}\delta ({\bf{r}}-{\bf{r}})\pa_{j}v_{i}({\bf{r'}})+ \pa_{i}\delta ({\bf{r}}-{\bf{r'}}),
\end{equation}
\begin{equation}
\label{ah}
\{v_{i}({\bf{r}}), v_{j}({\bf{r'}})\}=\frac{\pa_{j}v_{i}-\pa_{i}v{j}}{\rho}\delta ({\bf{r}}-{\bf{r'}})+\ta^{lm}\frac{\pa_{l}v_{i}\pa_{m}v_{j}}{\rho}\delta ({\bf{r}}-{\bf{r'}}).
\end{equation}
This is the complete NC algebra between the Eulerian fluid variables which reduces to the usual canonical form for $\theta_{ij}=0$.

\subsection{Algebraic consistency and Jacobi identity}

Consistency of any generalized Poisson structure in Hamiltonian framework demands validity of the Jacobi identity. In the present theory since we have posited a hitherto unknown algebra we must ensure that it satisfies the Jacobi identity.  Indeed, due to the non-linear nature of the NC algebra, explicit demonstration of Jacobi identity is quite involved. 

The Jacobi identity for a generic set of variables $a,b,c$, is given by,
$$J(a,b,c)=\{\{a,b\},c\}+\{\{b,c\},a\}+\{\{c,a\},b\}  =0.$$

  It proves to be convenient to work in momentum space  via Fourier transforms. We write down the density and  current in  momentum space as,
\begin{equation}
\label{ai}
\tilde{\rho}(\bf p)=\int d\bf r~ e^{i\bf p.\bf r}\rho({\bf r}),~~
\tilde{j^{i}(p)}=\int dr e^{ipr} j^{i}(r).
\end{equation}
 We recalculate the brackets in momentum space,
\begin{equation}
\nonumber
\{\tilde{\rho}({\bf{p}}), \tilde{\rho}({\bf{q}})\}=\{\int dr e^{i{\bf{p}}.{\bf{r}}}\rho({\bf{r}}), \int dr' e^{i{\bf{q}}.{\bf{r'}}}\rho({\bf{r'}})\}
\end{equation}
\begin{equation}
\label{al}
=- \ta^{ij}\int dr p_{j}(p_{i}+q_{i})e^{i(p+q)r}\rho({\bf{r}});~=i\hslash \ta^{ij}p_{i}q_{j}\tilde{\rho}({\bf{p}}+{\bf{q}})
\end{equation}
\begin{equation}
\label{am}
\{\tilde{\rho}({\bf{p}}),\tilde{j_{i}}({\bf{q}})\}=ip_{i}\tilde{\rho}({\bf{p}}+{\bf{q}})-\ta_{jk}q_{j}p_{k}\tilde{j_{i}}({\bf{p}}+{\bf{q}})
\end{equation}
To begin with, we take the $J(\rho , \rho , \rho)$ which in  momentum space reads,
\begin{equation}
	\nonumber
	J(\rho({\bf{p}}) , \rho ({\bf{q}}), \rho({\bf{r}}))=  [\theta^{ij}\theta^{lm}p^{i}q^{j}(p^{l}+q^{l})r^{m}]+ cyclic~ terms.
\end{equation}
After some algebra(details of the explicit demonstration of \eqref{yo1} are provided in the Appendix A2.) we recover
\begin{equation}
\label{yo1}
J(\rho({\bf{p}}),\rho({\bf{q}}),\rho({\bf{r}})) 
=\theta_{k}\theta_{n}\epsilon_{ijk}\epsilon_{lmn}[(p^{i}q^{j}(p^{l}+q^{l})r^{m}+q^{i}r^{j}(q^{l}+r^{l})p^{m}+r^{i}p^{j}(r^{l}+p^{l})q^{m}]=0.
\end{equation}
with,$$\theta_{ij}=\epsilon_{ijk}\theta_{k}.$$
To prove the next nontrivial identity (in   momentum space)  we need to check
\begin{equation}
J(\rho({\bf{p}}),\rho({\bf{q}}),v_{k}({\bf{r}}))=\{\{\rho({\bf{p}}),\rho({\bf{q}})\},v_{k}({\bf{r}})\}+\{\{\rho({\bf{q}}),v_{k}({\bf{r}})\},\rho({\bf{p}})\}+\{\{v_{k}({\bf{r}}),\rho({\bf{p}})\},\rho({\bf{q}})\}=0.
\end{equation}
Again a detailed but reasonably straightforward computation reveals that 
\begin{align}
\label{ak}
\nonumber
\{\{\rho({\bf{p}}),\rho({\bf{q}})\},v_{k}({\bf{r}})\}=\epsilon^{ijk}\theta_{k}p^{i}q^{j}[i(p_{k}+q_{k})\delta({\bf{p}}+{\bf{q}}+{\bf{r}})
+\epsilon_{lmn}\theta_{n}(p_{l}+q_{l})r_{m}v_{k}({\bf{p}}+{\bf{q}}+{\bf{r}})]=0.
\end{align}
which implies the validity of $J(\rho({\bf{p}}),\rho({\bf{q}}),v_{k}({\bf{r}}))=0$.

In a similar way, validity of rest of the non-trivial Jacobi identities can be checked as well.

\subsection{Modified Non Commutative algebra}
It is natural to consider further extensions of the
 NC structure that we have already considered by introducing  a new  set of NC  parameters $\sigma_{ij}$ in the
$\{\dot{X_{i}}({\bf{x}}), X_{j}({\bf{y}})\}$ bracket in \eqref{aa},
\begin{equation}
\label{2a}
\{X_{i}({\bf{x}}), X_{j}({\bf{y}})\} = \frac{\ta_{ij}}{\rho_{0}}\de ({\bf{x}}-{\bf{y}}),~
\{\dot{X_{i}}({\bf{x}}), X_{j}({\bf{y}})\}=\frac{1}{\rho_{0}}(\de_{ij} +\sigma_{ij})\de ({\bf{x}}-{\bf{y}}),~ \{\dot{X_{i}({\bf{x}})}, \dot{X_{j}}({\bf{y}})\}=0.
\end{equation}
Indeed, we emphasize that this new extension is not for purely academic purpose. We will see later that it has important consequence in cosmology.  Adopting the same procedure the new NC Euler algebra is found as,
\begin{equation}
\nonumber
\{\rho({\bf{r}}),\rho({\bf{r'}})\}= \rho_{0}^{2}\{\int dx \de(X({\bf{x}})-{\bf{r}}), 
\int dy \de(X({\bf{y}})-{\bf{r'}}\}
\end{equation} 
\begin{equation}
\label{2ae}
=-~\pa_{i}\rho \ta_{ij}\pa_{j}\de({\bf{r}}-{\bf{r'}}),
\end{equation}
\begin{equation}
\nonumber
\{\rho(r), j^{i}(r')\}= \rho_{0}^{2}\{\int dx \de(X({\bf{x}})-{\bf{r}}), 
\int dy~ \dot{X^{i}}(y) \de(X({\bf{y}})-{\bf{r'}}\}
\end{equation}
\begin{equation}
\label{2af}
=\rho({\bf{r'}})\pa_{i}\de({\bf{r}}-{\bf{r'}})
-\ta^{jk}\pa_{k}\de({\bf{r}}-{\bf{r'}})\pa_{j}j^{i}({\bf{r}})+\sigma_{ij}\rho({\bf{r'}})\pa_{j}
\de({\bf{r}}-{\bf{r'}})
\end{equation}
\begin{eqnarray}
\label{9q}
\{j^{i}({\bf{r}}),j^{k}({\bf{r}})\}= j_{k}({\bf{r}})\pa_{i}\de({\bf{r}}-{\bf{r'}})+j_{i}({\bf{r'}})\pa_{k}\de({\bf{r}}-{\bf{r'}})
+\sigma_{ij}j_{k}({\bf{r}})\pa_{j}\de({\bf{r}}-{\bf{r'}})\nonumber\\
+\sigma_{kj}j_{i}({\bf{r'}})\pa_{j}\de({\bf{r}}-{\bf{r'}})
+
\theta_{lm}\pa_{m}\de({\bf{r}}-{\bf{r'}})\pa_{l}\frac{j_{i}({\bf{r}})j_{k}({\bf{r}})}{\rho({\bf{r}})}
\end{eqnarray}
Again we provide the algebra between the velocity and the density,

\begin{equation}
\label{9w}
\{v^{i}({\bf{r}}), \rho({\bf{r'}})\}= \pa_{i}\de({\bf{r}}-{\bf{r'}})+\sigma_{ij}\pa_{j}
\de({\bf{r}}-{\bf{r'}}) +\theta_{kj}\pa_{k}\de({\bf{r}}-{\bf{r'}})\pa_{j}v_{i}({\bf{r'}})
\end{equation}
\begin{equation}
\label{2ah}
\{v_{i}({\bf{r}}), v_{j}({\bf{r'}})\}=\frac{\pa_{j}v_{i}-\pa_{i}v_{j}}{\rho}\de({\bf{r}}-{\bf{r'}})
+\ta^{lm}\frac{\pa_{l}v_{i} \pa_{m}v_{j}}{\rho}\de({\bf{r}}-{\bf{r'}})+ \frac{1}{\rho}(\sigma_{kj}\pa_{k}v_{i}-\sigma_{ik}\pa_{k}v_{j})
\de({\bf{r}}-{\bf{r'}}).
\end{equation}
Note that the  density algebra remains unaltered but the rest  receive $\sigma_{ij}$-contribution. Keeping the form of Hamiltonian unaltered,
$$H=\int dV ~ {\cal{H}}=\int (\frac{1}{2}\rho v^{2} +V(\rho))$$
  the NC-generalized  Euler dynamics follows,
\begin{equation}
\label{2an}
\dot{\rho}=\{\rho , H\}=-\pa_{i}(\rho v_{i})-\sigma_{ij}\pa_{j}(\rho v_{i})=-\pa_{i}(\rho v_{i} +\sigma_{ji}\rho v_{j}),
\end{equation}
\begin{equation}
\label{2al}
\dot{v_{k}}=\{v_{k}, H\}=-v_{i}\pa_{i}v_{k} - \sigma_{ij}v_{i}\pa_{j}v_{k}- \pa_{k}V'(\rho)- \sigma_{kj}\pa_{j}V'(\rho)+\theta_{ji}\pa_{i}V'(\rho)
\pa_{j}v_{k}.
\end{equation}
From \eqref{2an} it is clear that the total mass remains constant as the NC modification that showed up in the equation is a total derivative term. Though the flux changes, and both the NC terms have their contributions in it.

{\section {Noncommutative effects of fluid in Cosmology}}
Finally we discuss  the implications of  NC fluid dynamics in cosmological context. The present model lives in    flat space. (For introductory reference in cosmological perturbation see for example \cite{lyt}.)

 As is customary  in cosmology we now work in a comoving frame ($a(t), \bf x$)  where the map between laboratory and comoving coordinates ($\bf r$ and $a(t),\bf x$ respectively)  is given by,
 \begin{equation}
 \label{al}
 {\bf{r}}(t)=a(t){\bf{x}}
 \end{equation}
 with $a(t)$ being the scale factor and ${\bf{x}}$, the time independent  comoving distance. The canonical  continuity and Euler equations  in Friedmann-Robertson-Walker (FRW)  cosmology are given by.
\begin{eqnarray}
\dot{\rho}=-3H(\rho+P)=-3\frac{\dot{a}}{a}(\rho+P),
\label{nc14}
\end{eqnarray}
\begin{eqnarray}
\frac{\ddot{a}}{a}=-\frac{\rho+3P}{6M^2}+\frac{\Lambda}{3},
\label{nc15}
\end{eqnarray}
with   pressure $P $ and cosmological constant $\Lambda $ and $M=(8\pi G)^{-1/2}$ with $G$ the Newton's constant. $H(t)={\dot a}/a$ is the  Hubble parameter.  The  Friedmann equation follows:
\begin{eqnarray}
\frac{\dot{a}^2}{a^2}=H^2=\frac{\rho}{3M^2}+\frac{\Lambda}{3}-\frac{k}{a^2}.
\label{nc16}
\end{eqnarray}
It is well known that the canonical fluid equations in laboratory frame, when expressed in comoving frame, gets  mapped on to the above equations (\ref{nc14}, \ref{nc15}). Exploiting the same scheme we map the NC fluid equations (\ref{2an}, \ref{2al}) in comoving frame which will be interpreted as the NC FRW equations. 
\subsection{Cosmological perturbation}
Let us introduce the cosmological perturbation scheme. As usual we are assuming that at sufficiently large distance scales, (may be beyond galaxy clusters), the inhomogeneities average out leaving behind an isotropic and homogeneous background.

To that end velocities in the laboratory and comoving frames are related by,
\begin{equation}
\label{am1}
\dot{\bf{r}}={\bf{u}}=\dot{a}{\bf{x}}+ a\dot{\bf{x}}.
\end{equation}
Note that  ${\bf{x}}$ is now time-dependent and generates the  second term, known as peculiar velocity $a\dot{\bf{x}}={\bf v}$ which appears as a perturbation, (such that $|{
\bf v}|\ll |\dot{a}{\bf{x}}|$). Conventionally $\bf v$ is not considered in the canonical set of FRW equations. The relations between space and time derivatives between the laboratory and comoving frames are given by,
\begin{equation}
\label{aj4}
\frac{\pa}{\pa r}=\frac{1}{a}\frac{\pa}{\pa x},~~
\frac{\pa}{\pa t}|_{r}=\frac{\pa}{\pa t}|_{x}-\frac{\dot{a}}{a}(x.\pa_{x}).
\end{equation}

In constructing the perturbation theory it is customary to split the fields into a flat FRW background part and a perturbation part that can be analyzed order by order. However, we will find  (see (\ref{2bb1})) that the flatness condition is modified by NC contribution.

To  introduce perturbations one needs to perturb the metric in powers of a small parameter and subsequently expand the energy-momentum tensor in powers of the same parameter and compare the two sides of Einstein equation order by order. However, it is customary to absorb the small parameter in perturbations of the respective quantities and treat these as small with respect to the background \cite{perturb}. We adhere to the same convention in the following,
\begin{align}
\label{elm}
\nonumber
\rho({\bf{x}},t)=\rho_{0} (t)+\delta\rho(({\bf{x}},t))=\rho_{0}+\rho_{1}+\rho_{2}+....\\
\nonumber
P({\bf{x}},t)=P_{0}(t)+\delta P({\bf{x}},t)=P_{0}+P_{1}+P_{2}+...\\
\nonumber
H({\bf{x}},t)=H_{0}(t)+\delta H({\bf{x}},t)=H_{0}+H_{1}+H_{2}...\\
\nonumber
\phi({\bf{x}},t)=\phi_{0}(t)+\delta\phi({\bf{x}},t)=\phi_{0}+\phi_{1}+\phi_{2}+..\\
{\bf{u}}=\dot{a}{\bf{x}}+ {\bf{v}}=\dot{a}\bf{x}+v_{1}+v_{2}....
\end{align}
It needs to be stressed that this procedure of introducing inhomogeneity through perturbation about a homogeneous background is the conventional one. Keeping the background variables spatially invariant is valid since we are introducing the NC effect perturbatively and are considering NC corrections only up to  first order.   The novelty of our scheme lies in the fact that noncommutativity provides a natural seed for generating inhomogeneity.

  The peculiar velocity $\bf {v}$ in (\ref{elm}) is considered to be the perturbation in the velocity field.
  We here define a quantity namely density contrast(of order n) as, 
  \begin{equation}
  \label{yoi}
  \delta_{n}=\frac{\rho_{n}}{\rho_{0}}.
  \end{equation}
    $\phi $ is the gravitational potential defined by $$\nabla_{x}^{2}\phi=4\pi G a^{2}\rho . $$ Hence  the zero'th order background equation is,
  \begin{equation}
  \label{yt}
  \nabla_{x}^{2}\phi_{0}=4\pi G a^{2}\rho_{0}
  \end{equation}
  with the solution for the background potential
  \begin{equation}
  \label{yt1}
  \phi_{0}=\frac{2\pi}{3}G(ax)^{2}\rho_{0}
  \end{equation}
  using Newtonian model for gravitational potential due to a sphere of uniform density $\rho_0$. Furthermore, equations for the perturbations appear as  $$\nabla_{x}^{2}\phi_{n}=4\pi G a^{2}\rho_{n}$$
  where $n$ is the order of perturbation.
  \subsection{Noncommutative FRW from noncommutative fluid}
 In this section we will discuss the consequences of  noncommutative modified fluid from cosmological perspective. The crucial role played by $\sigma_{ij}$, appearing in our extended form of noncommutativity in fluid,  will come to the fore. 

The  first step will be to write the modified conservation equations \eqref{2an}, \eqref{2al},namely the continuity and the Euler equation, in comoving frame, exploiting \eqref{aj4}, 
\begin{equation}
\label{aj3}
\dot{\rho}+ 3\frac{\dot{a}}{a}\rho +\pa_{i}(\rho v_{i})+\frac{\sigma_{ij}}{a}\pa_{j}(\rho \dot{a}x_{i}+\rho v_{i})=0
\end{equation}
and,
\begin{eqnarray}
\nonumber
\ddot{a}x_{k} +\frac{\pa v_{k}}{\pa t}+\frac{\dot{a}}{a}
v_{k}+\frac{1}{a}v_{i}\pa_{i}v_{k}+\frac{\dot{a}}{a}\sigma_{ik}(\dot{a}x_{i}+v_{i})+\frac{1}{a}\sigma_{ij}(\dot{a}x_{i}+v_{i})\pa_{j}v_{k}\\=-\frac{1}{a}[\frac{\pa_{k}P}{\rho}+\sigma_{kj}\frac{\pa_{j}P}{\rho}+\frac{\dot{a}}{a\rho}\theta_{ik}\pa_{i}P+\frac{1}{a\rho}\theta_{ij}\pa_{i}P\pa_{j}v_{k}+\frac{4\pi}{3} a G\rho  x_{k}+\pa_{k}\phi] .
\label{ajb}
\end{eqnarray}
Note that $\partial_k P=\frac{\partial P}{\partial \rho}\partial_k \rho =c_s^2 \partial_k \rho$ where $c_s$ is the adiabatic sound speed. Thus the above equation reads
\begin{eqnarray}
\nonumber
\ddot{a}x_{k} +\frac{\pa v_{k}}{\pa t}+\frac{\dot{a}}{a}
v_{k}+\frac{1}{a}v_{i}\pa_{i}v_{k}+\frac{\dot{a}}{a}\sigma_{ik}(\dot{a}x_{i}+v_{i})+\frac{1}{a}\sigma_{ij}(\dot{a}x_{i}+v_{i})\pa_{j}v_{k}\\=-\frac{1}{a}[c_s^2\frac{\pa_{k}\rho}{\rho}+\sigma_{kj}\frac{\pa_{j}P}{\rho}+\frac{\dot{a}}{a\rho}\theta_{ik}\pa_{i}P+\frac{1}{a\rho}\theta_{ij}\pa_{i}P\pa_{j}v_{k}+\frac{4\pi}{3} a G\rho  x_{k}+\pa_{k}\phi] .
\label{ajb1}\end{eqnarray}

{\it{Continuity equation}}: Let us focus our attention on the Continuity equation \eqref{aj3}. If we expand \eqref{aj3} order by order, using \eqref{elm}, the background part satisfies
\begin{equation}
\label{2ao1}
\dot{\rho_{0}}+3\frac{\dot{a}}{a}\rho_{0}+\frac{1}{a}\sigma_{ij}\pa_{j}(\rho_{0}\dot{a}x_{i})=0 ~~\rightarrow \dot{\rho_{0}}+\frac{\dot{a}}{a}\rho_{0}(3+\sigma )=0
\end{equation}
where  $Tr(\sigma_{ij})=\sigma$. Clearly the NC effect  modifies  the background  continuity equation. If we set the NC contribution zero $(\sigma_{ij}=0)$ we will get the continuity equation \eqref{nc14} back (with zero pressure). {\footnote{ However, in an interesting variant of our model in \cite{kai} it is shown the $\theta_{ij}$ can also modify the continuity equation.}} We can make a further simplification by dropping the peculiar velocity contributions, namely $\bf{v}=0$ in the full equation (\ref{aj3}) leading  to 
\begin{equation}
\label{2ao}
\dot{\rho}+3\frac{\dot{a}}{a}\rho+\frac{1}{a}\sigma_{ij}\pa_{j}(\rho\dot{a}x_{i})=0.
\end{equation}
This is the usual way to generate the conventional ($\sigma_{ij}=\theta_{ij}=0$) FRW equation from fluid dynamics. In the present case we have derived the NC corrected continuity equation, even with vanishing   peculiar velocity.

{\it{Euler equation}}: Let us now concentrate  on the Euler equation \eqref{ajb}. We follow the conventional procedure of isolating  structurally similar terms in \eqref{ajb}  and requiring that the combinations vanish separately. In the present case the $x_i$-dependent terms read (with $\rho$ replaced by its homogeneous background value $\rho_0$):		
		\begin{equation}
		\label{3bam}
		[(\ddot{a}+\frac{4\pi}{3}  G\rho_{0})\delta_{ik}+ \dot{a} H\sigma_{ik}]x_{i} =0.
		\end{equation}
	To satisfy the above for arbitrary 	$x_i$ we require determinant of the coefficient matrix of $x_i$ to vanish,
		\begin{equation}
		\label{3acv}
		\begin{vmatrix}
		(\lambda +\dot{a}H\sigma_{11}) ~~ \dot{a}H \sigma_{12} ~~\dot{a}H\sigma_{13} \\
		\dot{a}H\sigma_{21} ~~ (\lambda + \dot{a}H\sigma_{22}) ~~ \sigma_{23}\dot{a}H \\
		\dot{a}H\sigma_{31} ~~  \dot{a}H\sigma_{32}  ~~(\lambda +\dot{a}H\sigma_{33})
		\end{vmatrix}
		=0 ,
		\end{equation}
				where, $$ \lambda =\ddot{a}+\frac{4\pi}{3}  G\rho_{0}.$$
	Expanding the determinant yields,	
		\begin{eqnarray}
		\nonumber
		(\lambda +\dot{a}H\sigma_{11})[(\lambda + \dot{a}H\sigma_{22})(\lambda +\dot{a}H\sigma_{33})-(\dot{a}H)^{2} \sigma_{23}\sigma_{32}] \\+
		(\dot{a}H)\sigma_{12}
		[(\dot{a}H)^{2}\sigma_{23}\sigma_{31}-\dot{a}H\sigma_{21}(\lambda +\dot{a}H\sigma_{33})]\\
		+ \dot{a}H\sigma_{13}[(\dot{a}H)^{2}\sigma_{21}\sigma_{32}-\dot{a}H\sigma_{31}(\lambda +\dot{a}H\sigma_{22})]=0.
		\label{3aqr}
		\end{eqnarray}
		Since we are interested in $O(\sigma )$ contributions, the above equation reduces to,
			\begin{equation}
		\label{3we}
		(\lambda)^{3}+ \lambda^{2}\dot{a}H(\sigma_{11}+\sigma_{22}+\sigma_{33})\approx 0,
		\end{equation}
		leading to
		\begin{equation}
		\label{3wf}
		\lambda+ \dot{a}H\sigma=0
		\end{equation}
		which is a   modified Euler equation in cosmology,
  \begin{equation}
\label{2bam}
\ddot{a}+\frac{4\pi}{3}  G\rho_{0}+ \dot{a} H\sigma =0,
\end{equation}
augmented by the $\sigma_{ij}$ contribution.

After a little more algebra we find that \eqref{2ao1} and \eqref{2bam} together yield,
\begin{equation}
\label{oun}
\frac{1}{2}\frac{d}{dt}(\dot{a}^{2})=\frac{4\pi G \rho_0}{3}[\frac{1}{\rho_0}(\frac{d}{dt}(\rho_0 a^{2})+ \frac{a}{\rho_0}\sigma\pa_{j}(\rho_0\dot{a}))]-\dot{a}^{2}H\sigma
\end{equation}
Finally the cherished  Friedmann
equation with NC correction  is recovered:
\begin{equation}
\nonumber
\frac{\dot{a}^{2}}{a^{2}}=\frac{8\pi G\rho}{3}-\frac{k}{a^{2}}+\frac{8\pi G}{3a^{2}}\int dt~ a (\rho_0 \dot{a})\sigma -\frac{2}{a^{2}}\int dt~ \dot{a}^{2}H\sigma .
\end{equation}

\begin{equation}
\label{2bb}
=\frac{8\pi G\rho}{3}-\frac{k_{eff}}{a^{2}},
\end{equation}
where 
\begin{equation}
\label{2bb1}
k_{eff}=k-\sigma (\frac{8\pi G}{3}\int dt~ a\dot{a} \rho_0 -{2} \int dt~ \dot{a}^{2}H ).
\end{equation}
The original (curvature)  constant $k$ is scaled to $0,~\pm 1$ signifying flat, closed or open   universe respectively. But in NC space this feature will be dictated by the effective curvature $k_{eff}$. For instance for a flat universe in NC cosmology $k_{eff}=0$ will lead to a relation,
\begin{equation}
\label{2ww}
k=\sigma (\frac{8\pi G}{3}\int dt~ a\dot{a} \rho_0 -{2} \int dt~ \dot{a}^{2}H )
\end{equation}

that can provide a bound on the value of $\sigma_{ij}$.

{\it{Cosmological perturbations}}: 

The aim of of introducing perturbations in the FRW "Standard Model" of cosmology  is to explain how large scale structures were formed in the
expanding Universe. In particular, this means that starting from an isotropic and
homogeneous universe with an average background density $\rho_0$, how does the fluctuation $\delta \rho =\rho-\rho_0$ grow so that the density contrast $\delta =\delta\rho /\rho_0$ can reach unity.   Once $\delta $ reaches values of the order of unity,  their growth becomes non-linear. From then onwards,  they rapidly evolve towards
bound structures such as star formation and other astrophysical process, eventually leading to  
formation of galaxies and clusters of galaxies.

Now, we would like to write the perturbation equation corresponding to the Euler equation \eqref{ajb} (without the terms in (\ref{2bam}) that has already been taken in to account). The perturbed equation is,
\begin{eqnarray}
\nonumber
\frac{\pa v_{k}}{\pa t}+(H_{0}+\delta H)
v_{k}+\frac{1}{a}v_{i}\pa_{i}v_{k}+(H_{0}+\delta H)\sigma_{ik}v_{i}+\frac{1}{a}v_{i}\pa_{i}v_{k}\\=-\frac{1}{a}[c_s^2\frac{\partial_k(\rho_0+\delta \rho)}{\rho_{0}+\delta\rho }+\sigma_{kj}\frac{\pa_{j}(P_{0}+\delta P)}{\rho_0+\delta\rho}+(H_{0}+\delta H)\theta_{ik}\frac{\pa_{i}(P_{0}+\delta P)}{\rho_{0}+\delta\rho }\nonumber
\\+\frac{1}{a}\theta_{ij}\frac{\pa_{i}(P_{0}+\delta P)}{\rho_{0}+\delta\rho }\pa_{j}v_{k}+\pa_{k}\de\phi].
\label{per}
\end{eqnarray}

 Here we will confine ourselves upto 1st order in perturbation so that terms of the form $\frac{\partial_k(\rho_0+\delta \rho)}{\rho_{0}+\delta\rho }\approx \frac{\partial_k \delta \rho}{\rho_{0}}$.  Thus we find

\begin{eqnarray}
\label{2bc}
\nonumber
\dot{v_{k}}^{1}+H_{0}(v_{k}^{1}+\sigma_{ik}v_{i}^{1})
=-[\frac{1}{a}c_s^2\frac{\pa_{k}\delta\rho}{\rho_{0}} +\pa_{k}\phi_{1}+\frac{1}{a\rho_{0}}H_{0}\theta_{ik}\pa_{i}P_{1}+ \frac{1}{a\rho_{0}}H_{1}\theta_{ik}\pa_{i}P_{0}\\ +\frac{1}{a\rho_{0}}\sigma_{kj}\pa_{j}P_{1}+
 \frac{1}{a^{2}\rho_{0}}\theta_{ij}\pa_{i}P_{0}\pa_{j}v_{k}]
\end{eqnarray} 
It is straightforward to see from (\ref{2ao1})  that the linear equations satisfied by the first order perturbations \cite{lyt}  are, 

\begin{equation}
\label{ccd}
\nonumber
 H^{1}=\frac{1}{3}\pa_{i}v_{i}^{1},~~~~~ \pa_{k}^{2}\Phi^{1}=4\pi G \de \rho^{1},
\end{equation}

\begin{equation}
\label{cce}
\dot{(\rho^{1})}=-\rho^{0} H^{1}(3+\sigma) -H^{0} \rho^{1}(3+\sigma) .
\end{equation}
Evidently the last relation is modified due to the non commutative modifications in \eqref{2a}.
Here we recall that $\rho_{0}\propto a^{-3}$ which leads to a further simplification \cite{lyt} in the last relation in \eqref{cce},
\begin{equation}
\label{yooo}
\dot{(\de^{1})}=- H^{1}(3+\sigma).
\end{equation}

We are interested in finding out the changes brought in by the non commutative considerations in the density perturbation equation.
For that we would like to work with the density contrast \eqref{yoi} over $\de \rho$ and derive  the density perturbation equation. Taking  divergence of the perturbation equation \eqref{per} results in,
\begin{equation}
\label{ca}
\pa_{k}\dot{v_{k}}^{1}+H_{0}\pa_{k}(v_{k}^{1}+\sigma_{ik}v_{i}^{1})=-\frac{1}{a}[c_s^2\frac{\pa_{k}^{2}\rho^1}{\rho_{0}}+\pa_{k}^{2}\phi^{1}+\sigma_{kj}\frac{\pa_{k}\pa_{j}P^{1}}{\rho_{0}}+\frac{1}{a^{2}\rho_{0}}\theta_{ij}\pa_{i}P_{0}\pa_{j}v_{k}].
\end{equation}
Now, the relation connecting the divergence of the  peculiar velocity and the Hubble parameter \eqref{cce},  is used. Some more algebra yields,

\begin{equation}
\label{cc}
\dot H^{1}=-2H^{0} H^{1}-\frac{1}{3a}[c_s^2\frac{\pa_{k}^{2}\rho^{1}}{a\rho_{0}}+\frac{\pa_{k}^{2}\phi^{1}}{a}+H_{0}\sigma_{ik}\pa_{k}v_{i}^{1}+\sigma_{kj}\frac{\pa_{k}\pa_{j}P^{1}}{a\rho_{0}}+\frac{1}{a^{2}\rho_{0}}\theta_{ij}\pa_{i}P_{0}\pa_{j}v_{k}].
\end{equation}

\subsection { Wave Equation for Growth
of Small Density Perturbations}
 Eventually using \eqref{yooo}  we derive the cherished form  density perturbation equation:
\begin{equation}
\label{cd}
\ddot{\delta^1}=-2H_{0}\dot{\delta^1}+\frac{(3+\sigma )}{3a}[H_{0}\sigma_{ik}\pa_{k}v_{i}^{1}+c_s^2\frac{\pa_{k}^{2}\delta^{1}}{a}+\sigma_{kj}\frac{\pa_{k}\pa_{j}P^{1}}{a\rho_{0}}
+\frac{\pa_{k}^{2}\phi^{1}}{a}+\frac{1}{a^{2}\rho_{0}}\theta_{ij}\pa_{i}P_{0}\pa_{j}v_{k}].
\end{equation}

 The noncommutative parameter $\sigma_{ij}$  being  small, we can ignore terms  quadratic in $\sigma_{ij}$. The term containing $\theta_{ij}$ is ignored compared to the other terms since it varies as $\frac{1}{a^3}$. Furthermore, seeking solutions of the form $\delta^1 \sim exp~i(\bf{k_c}.\bf{x}-\omega t)$ we note that  $c_s^2\pa_{k}^{2}\delta^{1}=-c_s^2k_c^2\delta^{1}=-c_s^2k^2a^2\delta^{1}$ where $\bf{k}_c$ and $\bf{k}$ are respectively the comoving and proper wave vector,
 \begin{eqnarray}
 \label{hast}
 \nonumber
 \ddot{\delta^1}=-2H_{0}\dot{\delta^1}+
  \frac{\pa_{k}^{2}\phi^{1}}{a^{2}}+c_s^2\frac{\pa_{k}^{2}\delta^{1}}{a^2}+\frac{\sigma }{3}\frac{\pa_{k}^{2}\phi^{1}}{a^{2}}+\frac{1}{a} H_{0}\sigma_{ik}\pa_{k}v_{i}^{1}+\sigma_{kj}\frac{\pa_{k}\pa_{j}P^{1}}{a^2\rho_{0}}\\
 =-2H\dot{\delta^1}+
 ( 4\pi G\rho_{0}-c_s^2{k}^{2})\de ^{1} +\frac{4\pi G \rho_{0}}{3}\sigma \de^{1} +\frac{1}{a}\sigma_{ik}( H_{0}\pa_{k}v_{i}^{1}+\frac{\pa_{i}\pa_{k}P^{1}}{a\rho_{0}}).
 \end{eqnarray} 
 
  Finally we have reached at our goal of obtaining the density perturbation equation. This equation govern the dynamics of small density fluctuations in a noncommutative fluid for an expanding background cosmology without cosmological constant.  
  
  We rewrite the above equation in the convenient form,
  
  \begin{eqnarray}
  \label{hast1}
    \ddot{\delta^1}
  =-2H\dot{\delta^1}+
   4\pi G\rho_{0}(1+\frac{\sigma}{3})\de ^{1}-c_s^2{k}^{2}\de ^{1} +\Sigma, 
  \end{eqnarray} 
  where $\Sigma =\sigma_{ik}  \frac{\pa_{i}\pa_{k}P^{1}}{a^2\rho_{0}}$ where we have dropped the term  $\frac{1}{a}\sigma_{ik} H_{0}\pa_{k}v_{i}^{1}$ from $\Sigma $ since it is $O(\sigma v^1)$.

   Thus $\sigma$ and $\Sigma$ are both NC contributions.
  
  {\it{Jeans' instability in expanding medium}}: The pressure terms are negligible except on small scales just before the matter radiation equality \cite{lyt}. Hence in the long wavelength limit{\footnote{Long wavelength limit refers to $\lambda\rangle\rangle \lambda_J=c_s\sqrt{\frac{\pi}{G\rho_0}}$, $\lambda_j$ is the Jeans' wavelength in conventional cosmology.}}  we can drop the terms generated by pressure and consider a reduced form of (\ref{hast1}),
\begin{eqnarray}
 \label{hast}
 \ddot{\delta^1}=-2H_{0}\dot{\delta^1}+
  4\pi G\rho_{0}(1+\frac{\sigma}{3})\de ^{1} .
 \end{eqnarray}

  In the linear regime, density fluctuations on different scales evolve independently. Thus it is useful to write the equations \eqref{yooo}, \eqref{hast1} in the Fourier space as,
   $$H_{k}^{1}=-\frac{\dot{\de_{k}}}{3+\sigma},$$ 
  \begin{equation}
 \label{yaa}
 \ddot{\de_{k}^{1}}+2H_{0}\dot{\de_{k}^{1}}=4\pi G\rho_{0}(1+\frac{\sigma}{3})\de_{k}^{1} + \Sigma_{k}
 \end{equation}
where $\Sigma_{k}= -\sigma_{ik}\frac{k^2}{a^2\rho_{0}}P^{1}_{k}$ is the $\Sigma$ written in the Fourier space. We will drop this term since we are neglecting pressure as explained earlier. The modified \eqref{yaa} can be written as,
\begin{equation}
\label{pa2}
\ddot{\de_{k}^{1}}+2H_{0}\dot{\de_{k}^{1}}=4\pi G\rho_{0}(1+\frac{\sigma}{3})\de_{k}^{1}.
\end{equation}
We will try to find solution of the equation \eqref{pa2} in a flat space which implies at critical density $(\rho=\rho_{c})$. Under these conditions we have to find out the dependence of $a$ and $\rho_{0}$ on time and subsequently we would like to solve \eqref{pa2}. 

Before proceeding further to derive explicit form of $\delta_k^1$ it is important to stress that the background, (about which the fluctuations are being studied), is no longer the conventional one since it has already received a NC correction, as is seen from (\ref{2ao1}) {\footnote{We thank the referee for pointing this out.}}. So the first task is to ascertain the NC modified background density $\rho_0$ for which we  consider the modified background continuity equation \eqref{2ao1}. The  solution is given by,
 \begin{equation}
 \label{n1}
\rho_{0}=\bar{\rho} a^{-(3+\sigma)}.
\end{equation}
 As we are confining ourselves upto first order in $\sigma$ we are allowed to use the canonical time dependence of  $a(=A_{0}t^{\frac{2}{3}})$\cite{lyt} and the solution of the modified continuity equation (\ref{2ao1}) to get the time dependence of $k$ under flat space condition from \eqref{2ww}. A straightforward computation yields\footnote{$k=\sigma (\frac{8\pi G}{3}\bar{\rho}\int dt~ a\dot{a} a^{-(3+\sigma)} -{2} \int dt~ \dot{a}^{2}H )$},
\begin{equation}
\label{la1}
k(t)=\frac{8}{3}\sigma t^{-2/3}(-\pi G\bar{\rho}A_{0}^{-(1+\sigma)} t^{-2\sigma /3}+\frac{A_{0}^{2}}{3}).
\end{equation}
Quite obviously this $k(t)$ is proportional to the NC parameter $\sigma $ and vanishes in the conventional (flat space) case. On using this $k$  in the Friedmann equation \eqref{nc16} (with $\Lambda =0$, no cosmological constant) we get,
\begin{eqnarray}
\frac{\dot{a}^2}{a^2}=H^2=\frac{\rho_0}{3M^2}-\frac{\frac{8}{3}\sigma t^{-2/3}(-\pi G\bar{\rho}A_{0}^{-(1+\sigma)} t^{-2\sigma /3}+\frac{A_{0}^{2}}{3})}{a^2}.
\label{nc21}
\end{eqnarray}
We want to obtain the solution of $a$ as a polynomial in $t$ restricting ourselves to the first non-trivial $\sigma$-correction. In the RHS of (\ref{nc21}) we substitute 
\begin{equation}
\label{la100}
\rho_0=\bar\rho a^{-(3+\sigma)},~~a=A_{0}t^{2/3},
\end{equation}
that amounts to taking account of the $\sigma$-corrected background and conventional form of $a(t)$ so that (\ref{nc21}) will yield the $O(\sigma)$ corrected $a(t)$. Here $A_{0}$ and $\bar\rho$ are simply  two constants that take care of the dimensions. It is straightforward to get a solution of the form,
\begin{equation}
\label{vae}
t=Aa^{\frac{3+\sigma}{2}}+Ba^{3(\frac{1+\sigma}{2})}
\end{equation}
where  $A$ and $B$ are constants, $$A=\frac{2(1-\sigma)}{3+\sigma}\sqrt{\frac{3}{8\pi G\bar\rho}},~~B=\frac{8\sigma A_{0}^{3}}{27(1+\sigma)}(\frac{3}{8\pi G\bar\rho})^{\frac{3}{2}}.$$
 We need to invert (\ref{vae}) to express $a$ as a function of $t$ in the familiar form,
\begin{equation}
\label{vyo}
a=(\frac{t}{A})^{\frac{2}{3+\sigma}}[1-\frac{BA_{0}^{2\sigma /3}}{A}t^{\frac{2\sigma}{3}}]^{\frac{2}{3+\sigma}}
\end{equation}
where, $\frac{B}{A}=\frac{2A_{0}^{3}\sigma}{3\pi G\bar\rho}$. First of all it is reassuring to note that for $\sigma =0$ the familiar form, $a(t)\sim t^{2/3}$ is recovered. For convenience we further approximate $a(t)\sim t^{2/(3+\sigma )}$ in subsequent analysis. Putting everything together in (\ref{pa2})  provides the cherished evolution equation of $\delta_k^1$:
\begin{equation}
 \label{yaq}
\ddot{\de_{k}^{1}}+\frac{4(1-\sigma /3)}{3t}\dot{\de_{k}^{1}}-\frac{2}{3t^2}(1+\frac{\sigma}{6})\de_{k}^{1}=0.
 \end{equation}
 By inspection a power law solution  $\de_k^1 \sim t^n$ yields
 \begin{equation}
 	\label{solu}
 	n=\frac{1}{6}[-1 + \frac{4\sigma}{3} \pm 5\sqrt{1-\frac{11}{75}\sigma}]\approx   \frac{1}{6}[-1 + \frac{4\sigma}{3} \pm 5(1-\frac{11}{150}\sigma)].
 \end{equation}
 The NC corrected values of $n$ are
\begin{equation}
	\label{solu1}
n=\frac{2}{3}+\frac{29}{180}\sigma  ,~~n=-1+\frac{51}{180}\sigma .
\end{equation}
Note that $\sigma$ can be either positive or negative.
Positive and negative values of $n$ signify growing or decaying modes.	
Obviously allowed values of $\sigma $ have to be such that the original nature of the mode (growing or decaying) is not altered.
 This constitutes the other significant result of our paper. In the next section we discuss some of the consequences of NC fluid model in cosmology.
  \section{Noncommutative effect on Hubble parameter and linear growth of structure}
  Indeed it is pertinent to ask to what extent can NC  affect the curvature and related evolutionary history of the universe in a quantitative way. Generically numerical upper bounds of NC parameters, obtained from areas in quantum mechanics or particle physics are in fact extremely small. From a theoretical perspective NC effects are expected to become relevant at approximately around Planck scale when the spacetime continuum  tends to get replaced by discreteness with noncommutativity manifesting itself by inducing  an inherent length scale. However, we should emphasize the distinction between the above scenario and the present context because,  strictly speaking, in the latter,  we are dealing with a {\it{non-canonical}}   Poisson bracket structure  in classical physics, rather than a noncommutative structure in the quantum commutators. Even though the non-canonical structure carries the legacy of the NC-extended (Heisenberg) quantum commutation relations or vice-versa and both affect classical and quantum physics respectively in similar fashion, there are important differences between NC generalizations in Poisson brackets in classical mechanics and commutation relations in quantum mechanics, notable among them being that dimensionally the NC parameters in the two scenarios are different.  
  
  One of the most important observables in cosmology is the Hubble parameter $H(t)$. Let us concentrate on the NC effect on $H$.
    Using the explicit form of NC-modified scale factor $a(t)$ we compute $H(t)$ and plot it against $t$ for two values of $\sigma =\pm 0.1$ and $\sigma =\pm 0.5$ (since $\sigma$ can take positive or negative values). This is depicted in Figure 1 where profiles for $H(t)$ for $\sigma =\pm 0.1$ and  $\sigma =\pm 0.5$ are plotted. These can be compared with the conventional case,  $\sigma =0$, the middle  black line. In our simplified scheme we have 
  \begin{equation}
  \label{h}
  H(t)=\frac{2}{(3+\sigma)t}.
    \end{equation}
    Thus larger negative values of $\sigma$ tend to stay more and more above the $\sigma =0$ line whereas larger positive values of $\sigma$ stay below the $\sigma =0$ line. Comparing with a conventional matter dominated universe $H\sim 2/(3t)$, one might conclude that the NC correction for positive    $\sigma$ reduces $H$ indicating that the rate of expansion of universe slows down, thereby   simulating a dark matter like behavior whereas values of negative $\sigma$ seem to behave in a way that opposes the conventional matter contribution. Furthermore  Hubble parameter also indicates the physical distance at which objects are receding at the speed of light, which is referred to as the Hubble distance given by $R_H=c/H$. Thus the Hubble distance increases (decreases) for negative (positive) values of $\sigma $.
    
  \begin{figure}
  	\begin{center}
  		\includegraphics[width=0.45\textwidth]{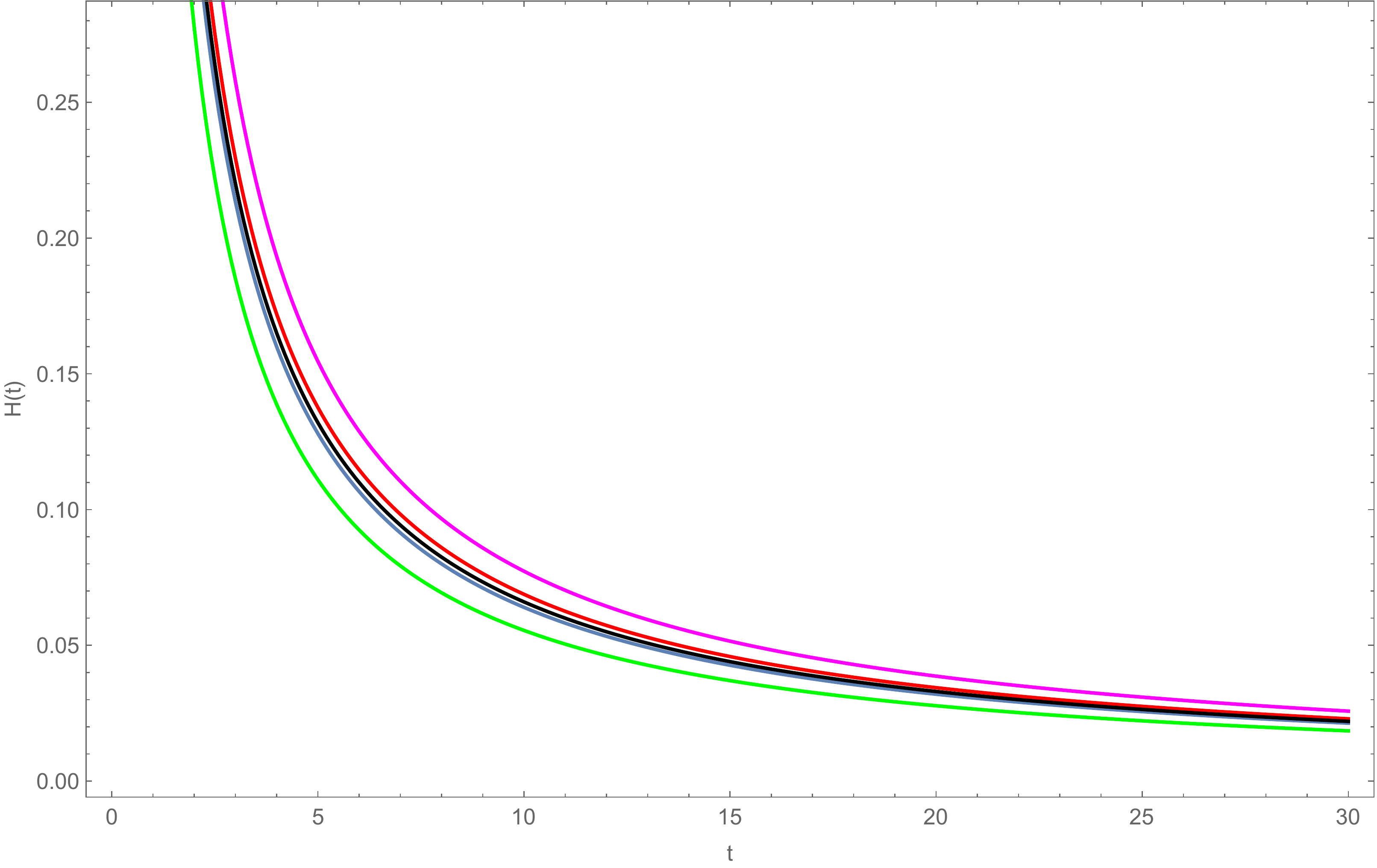}
  		\caption{$H(t)$  is plotted against $t$  for $\sigma =0$ black line (conventional case), $\sigma =\mp 0.1$  blue and green lines respectively and $\sigma =\mp 0.5$  pink and red lines respectively.  }
  		\label{fig-p-vs-H}
  	\end{center}
  \end{figure}
  
  The other object of interest related to structure formation is the NC-correction in the evolution  of the density contrast modes $\delta_k^1\sim t^n$ where NC-modified $n$ is provided in (\ref{solu}). Once again for $\sigma =0$ the conventional values $n=-1$ and $n=+2/3\approx 0.66$ are recovered out of which the latter increasing mode is of interest. From (\ref{solu1}) we get for $\sigma =\pm 0.1$, $n$ changes to $0.68,~0.63$ respectively   and for  $\sigma =\pm 0.5$, $n$ changes to $0.74,~0.58$ respectively   for the increasing mode. In Figure 2 we have plotted $\delta_k^1$ against $t$ for the above four values of $n$ along with $n=+2/3$ (for $\sigma = 0$) for comparison. The nature of the profiles presented in Figure 2 reveal that positive values of $\sigma$ enhances the growing modes so that structure formation is favored. In this sense our model of generalized fluid dynamics in the cosmological perspective becomes interesting since it might lead to a dark matter model, (that is essential for explaining the observed large-scale structure in the Universe), remaining rooted in classical physics.
   \begin{figure}
 	\begin{center}
 		\includegraphics[width=0.45\textwidth]{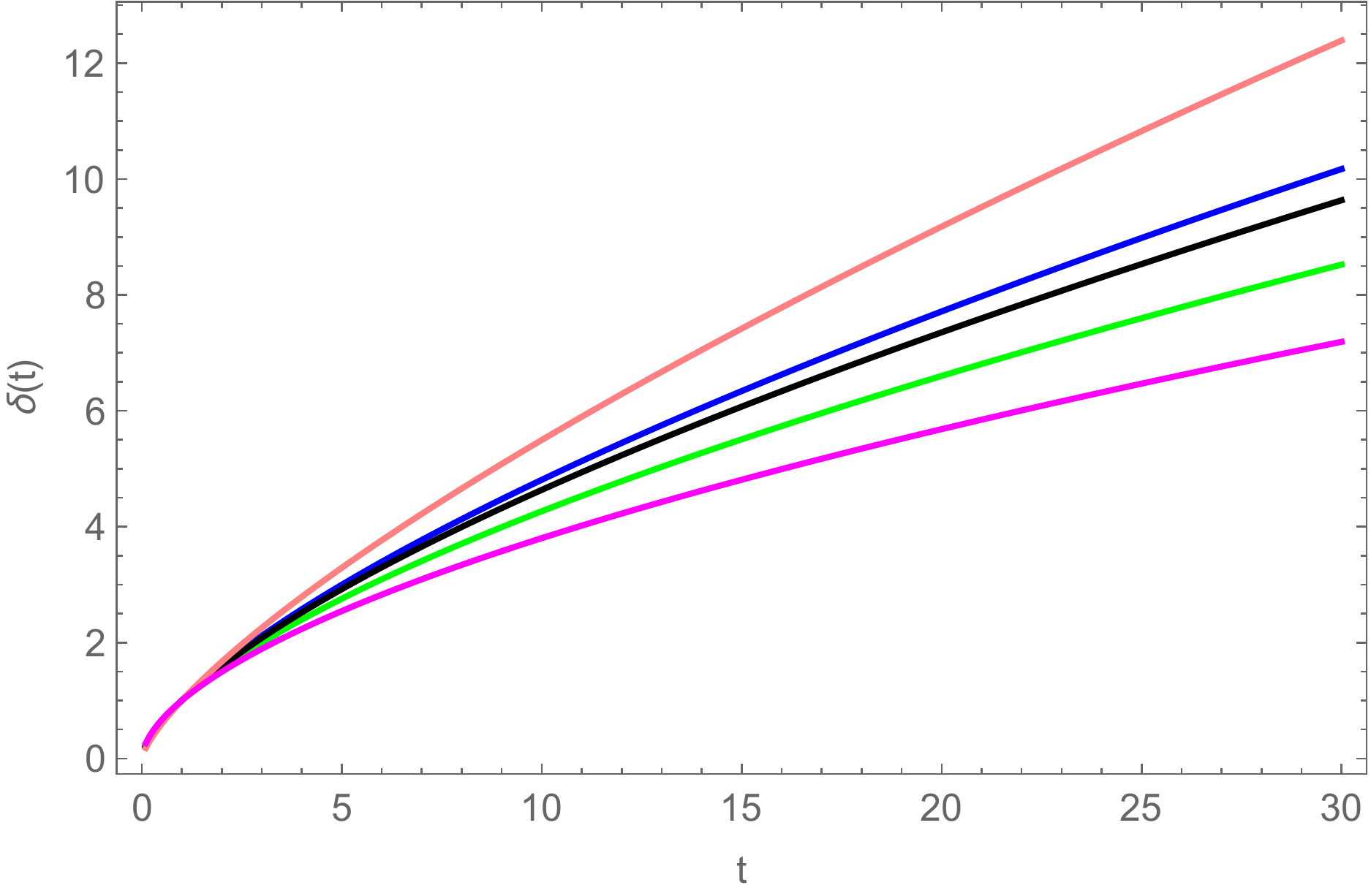}
 		\caption{$\delta_k^1(t)\sim t^n$  is plotted against $t$  for $n=2/3,~\sigma =0$ black line (conventional case), $n=0.68,~ 0.63$; ($\sigma =\pm 0.1$) for blue and green lines respectively and  $n=0.74,~ 0.58$; ($\sigma =\pm 0.5$) for orange and magenta lines respectively.}
 		\label{fig-p-vs-H}
 	\end{center}
 \end{figure}

  \section{Conclusion}
 In the present paper we have considered cosmological implications of a generalized fluid model in non-relativistic Newtonian framework. Our model is a non-trivial extension of  noncommutative fluid model, recently proposed by our groups. In the first part we rigorously derive formal aspects of the noncommutative  Hamiltonian fluid model. In particular we clarify issues related to the Jacobi identity of the NC fluid variable algebra. This NC algebra leads to a modified form of dynamical fluid equations, {\it{i.e.}} continuity and Euler equations. These are the starting point of the second part of our work where we discuss cosmological effects of the noncommutative extension. This constitute the major part of the paper.
 
 In the second part we introduce cosmological perturbations and explicitly show how the behavior of growing and decaying modes of density contrast are affected by noncommutative (or non-canonical, which is probably more appropriate as pointed out in the paper) corrections. We have explicitly demonstrated that the positive or negative values of the noncommutative parameter $\sigma$ can decrease or increase  the Hubble parameter respectively. The former can be identified with an effective model for dark matter. Similarly positive $\sigma$ enhances the increasing mode of density contrast which also agrees with the dark matter interpretation mentioned above.
 
  We have considered the simplest form of approximation and a more detailed analysis of the model is needed. Specifically one of our future projects is to find solutions of the scale factor directly computed from the noncommutativity extended equations derived here. Finally it would be interesting to exploit the rigorous cosmological averaging principles developed by Buchert and coworkers \cite{buchert} in the present context where the modifications stem from the fact that the evolution and averaging of dynamical variables do not commute.

\section*{Appendix}
{\bf{A. Calculation of the brackets between $\rho$ and $\bf{j}$}}
\begin{equation}
\nonumber
\{\rho(r), j^{i}(r')\}= \rho_{0}^{2}\{\int dx \de(X(x)-r), 
\int dy~ \dot{X^{i}}(y) \de(X(y)-r')\}
\end{equation}
\begin{equation}
\nonumber
=\rho_{0}^{2}[\int dx dy \{\de(X(x)-r),\dot{X^{i}}(y)\}\de(X(y)-r')+ \{\de(X(x)-r), \de(X(y)-r')\}\dot{X^{i}}(y)]
\end{equation}
\begin{equation}
\label{af}
=\rho(r')\pa_{i}\de(r-r')-\ta^{jk}\pa_{k}\de(r-r')\pa_{j}j^{i}(r)
\end{equation}

{\bf{B. Explicit calculation of one of the Jacobi identities,}}
\begin{equation}
\nonumber
J(\rho(p),\rho(q),\rho(r))=\{\{\rho(p),\rho(q)\},\rho(r)\} +cyclic terms
\end{equation}
\begin{equation}
\nonumber
=\theta_{k}\theta_{n}\epsilon_{ijk}\epsilon_{lmn}[(p^{i}q^{j}(p^{l}+q^{l})r^{m}+q^{i}r^{j}(q^{l}+r^{l})p^{m}+r^{i}p^{j}(r^{l}+p^{l})q^{m}]
\end{equation}
\begin{align}
\nonumber
=\theta_{k}\theta_{n}[\de_{il}(\de_{jm}\de_{kn}-\de_{jn}\de_{km})-\de_{im}(\de_{jl}\de_{kn}-\de_{jn}\de_{kl})+\de_{in}(\de_{jl}\de_{km}-\de_{jm}\de_{kl})][(p^{i}q^{j}(p^{l}+q^{l})r^{m}+\\
q^{i}r^{j}(q^{l}+r^{l})p^{m}+r^{i}p^{j}(r^{l}+p^{l})q^{m}]
\end{align}
\begin{align}
\nonumber
=\theta^{2}(p^{2}(q.r)+(p.q)(q.r)-(p.r)(p.q)-q^{2}(p.r))-\theta_{m}\theta_{n}(p^{2}q^{n}r^{m}+(p.q)q^{n}r^{m}-(p.q)p^{n}r^{m}\\
\nonumber
-q^{2}p^{n}r^{m})+\theta_{n}\theta_{l}((p.r)q^{n}p^{l}+(p.r)q^{n}q^{l}-p^{n}p^{l}(q.r)-(q.r)p^{n}q^{l})\\
\nonumber
+\theta^{2}(q^{2}(r.p)+(q.r)(r.p)-(q.p)(q.r)-r^{2}(q.p))-\theta_{m}\theta_{n}(q^{2}r^{n}p^{m}+(q.r)r^{n}p^{m}-(q.r)q^{n}p^{m}\\
-r^{2}q^{n}p^{m})+\theta_{n}\theta_{l}((q.p)r^{n}q^{l}+(q.p)r^{n}r^{l}-q^{n}q^{l}(r.p)-(r.p)q^{n}r^{l})\\
\nonumber
+\theta^{2}(r^{2}(p.q)+(r.p)(p.q)-(r.q)(r.p)-p^{2}(r.q))-\theta_{m}\theta_{n}(r^{2}p^{n}q^{m}+(r.p)p^{n}q^{m}-(r.p)r^{n}q^{m}\\
\nonumber
-p^{2}r^{n}q^{m})+\theta_{n}\theta_{l}((r.q)p^{n}r^{l}+(r.q)p^{n}p^{l}-r^{n}r^{l}(p.q)-(p.q)r^{n}p^{l})
\end{align}

\begin{align}
\nonumber
=-\theta_{m}\theta_{n}(p^{2}q^{n}r^{m}+(p.q)q^{n}r^{m}-(p.q)p^{n}r^{m}
-q^{2}p^{n}r^{m})+\theta_{n}\theta_{l}((p.r)q^{n}p^{l}-(q.r)p^{n}q^{l})\\
\nonumber
-\theta_{m}\theta_{n}(q^{2}r^{n}p^{m}+(q.r)r^{n}p^{m}-(q.r)q^{n}p^{m}
-r^{2}q^{n}p^{m})+\theta_{n}\theta_{l}((q.p)r^{n}q^{l}-(r.p)q^{n}r^{l})\\
\nonumber
-\theta_{m}\theta_{n}(r^{2}p^{n}q^{m}+(r.p)p^{n}q^{m}-(r.p)r^{n}q^{m}
-p^{2}r^{n}q^{m})+\theta_{n}\theta_{l}((r.q)p^{n}r^{l}-(p.q)r^{n}p^{l})\\
=0
\end{align}

{\bf{C. Explicit calculation of the $\rho$, {\bf{j}} with the modified NC algebra}}
\begin{eqnarray}
\nonumber
\{\rho (r), j^{i}(r')\}=\rho_{0}^{2}\int dx dx' \{\de(X(x)-r), \dot{X^{i}}\de(X(x')-r')\}\\
\nonumber
=\rho_{0}^{2}\int dx dx'\int [\pa_{j}^{X(x)}\de((X(x)-r))\pa_{k}^{X(x')}\de((X(x')-r'))\{X_{j},X_{k}\}\dot{X_{i}}(x')\\
\nonumber
+ \pa_{j}^{X(x)}\de((X(x)-r))\de(X(x')-r')\{X_{j},\dot{X_{i}\}}]\\
\nonumber
=\rho_{0}\int dx dx'\int[\pa_{j}^{X(x)}\de((X(x)-r))\pa_{k}^{X(x')}\de((X(x')-r'))\theta_{jk}\de(x-x')\dot{X_{i}}(x')\\
\nonumber
-\pa_{j}^{X(x)}\de((X(x)-r))\de(X(x')-r')(\de_{ij}+\sigma_{ji})\de(x-x')]\\
\nonumber
=-\rho_{0}\theta_{jk}\pa_{k}^{r}\de(r-r')\pa_{j}^{r}[\int dx \dot{X_{i}}\de(X(x)-r)]-\pa_{i}^{r'}\de(r-r')\rho(r')-\sigma_{ji}\pa_{j}^{r'}\de(r-r')\rho(r')\\
=\rho(r')\pa_{i}\de(r-r')-\ta^{jk}\pa_{k}\de(r-r')\pa_{j}j^{i}(r)+\sigma_{ji}\rho(r')\pa_{j}\de(r-r')
\end{eqnarray}


\begin{thebibliography}{99}
	\bibitem{ar} 	R. Banerjee, S. Ghosh, A. K. Mitra;    EPJC (2015) 75:207;
	
	A. K. Mitra, R. Banerjee, S. Ghosh;  International Journal of Modern Physics A Vol. 32, No. 36, 1750210 (2017).
	
	 \bibitem{pra1}	P. Das, S. Ghosh, 	10.1140/epjc/s10052-016-4488-8, 		  
	 	
	 	 P. Das, S. Ghosh; Back Reaction Inhomogeneities in Cosmological Parameter Evolution via Noncommutative Fluid,
	 	 arXiv:1804.07475 ; 	
		Praloy Das, Subir Ghosh
	 Phys. Rev. D 96, 111901 (2017)
	 
	 \bibitem{kai} Kai Ma, arXiv:1801.02533 ,
	 Fluid Dynamics on Noncommutative Space
	 
	 \bibitem{van} M. C. B. Abdalla, L. Holender, M. A. Santos, and I. V. Vancea
	 Phys. Rev. D 86, 045019 (2012); L. Holender, M. A. Santos, M. T. D. Orlando, and I. V. Vancea
	 Phys. Rev. D 84, 105024 (2011); 
	  M.V. Marcial, A.C.R. Mendes, C. Neves, W. Oliveira and F.I. Takakura, Phys. Lett. A 374 (2010) 3608.
	\bibitem{ncrev}N. Seiberg, E. Witten, JHEP 9909 032 (1999) [hep-th/9908142].; For reviews see M.R.Douglas and N.A.Nekrasov, Rev. Mod. Phys. 73 977 (2001) [hep-th/0106048]; R. J. Szabo, Phys.
	Rep. 378 207 (2003) [hep-th/0109162]; R. Banerjee, B. Chakraborty, S. Ghosh, P. Mukherjee, S.
	Samanta Found.Phys.39:1297-1345, 2009 .
	\bibitem{jac}  R. Jackiw, V.P. Nair, S.-Y. Pi, A.P. Polychronakos,
	J.Phys. A37 (2004) R327-R432 	(arXiv:hep-ph/0407101).
		\bibitem{mal}J. Maldacena et al., Phys. Rept. 323 (2000) 183; 
	S. S. Gubser, Phys. Rev. D 78 (2008) 065034; 
	S. A. Hartnoll, C. P. Herzog and G. T. Horowitz, Phys. Rev. Lett. 101 (2008) 031601.
		\bibitem{gg} S. Bhattacharyya, V. E. Hubeny, S. Minwalla, and M. Rangamani, JHEP 02 (2008) 045; S. Bhattacharyya, V. E. Hubeny, R. Loganayagam, G. Mandal, S. Minwalla, T. Morita,
	M. Rangamani, and H. S. Reall, JHEP 2008 (2008), no. 06 055. See for example N. Ambrosetti, J. Charbonneau, S. Weinfurtner, 
	The fluid/gravity correspondence:
	Lectures notes from the 2008 Summer School on Particles, Fields, and Strings, UBC, Canada.
	 \bibitem{sny} H. S. Snyder, Phys. Rev. 71 38 (1947); ibid 72 68 (1947); See also C. N. Yang, Phys. Rev. 72 874 (1947) .
	 
	 \bibitem{dirac} P.A.M.Dirac, Lectures on Quantum Mechanics, Yeshiva University Press, New York, 1964.
	 
\bibitem{lyt} A. R. Liddle, D. H. Lyth,
Cambridge University Press, 13-Apr-2000.
\bibitem{perturb} R. Durrer 2. Cosmological Perturbation Theory, in, E. Papantonopoulos (eds) The Physics of the Early Universe. Lecture Notes in Physics, {{653}}. Springer, Berlin, Heidelberg;

   P. Peter 'Cosmological Perturbation Theory',
   arXiv:1303.2509 [astro-ph.CO]
   



\bibitem{buchert}  T.Buchert, M.Kerscher and C.Sicka, Phys. Rev. D 62, 043525(2000); 
T. Buchert and J. Ehlers, Astron. Astrophys.	320,1 (1997);  T. Buchert, Gen. Relativ. Gravit.	32, 105 (2000); Astron. Astrophys.223,9 (1989);  ASP Conference Series (1996) 349-356	(astro-ph/9512107); Gen.Rel.Grav. 32 (2000) 105-125
(gr-qc/9906015); Gen.Rel.Grav.33:1381-1405,2001 (gr-qc/0102049).

\end{thebibliography}
\end{document}